# A general mixture equation of state for double bonding carboxylic acids with ≥ 2 association sites

Bennett D. Marshall

*ExxonMobil Research and Engineering, 22777 Springwoods Village Parkway, Spring TX 77389 USA*

**Abstract**

In this paper we obtain the first general multi-component solution to Wertheim's thermodynamic perturbation theory for the case that molecules can participate in cyclic double bonds. In contrast to previous authors, we do not restrict double bonding molecules to a 2-site association scheme. Each molecule in a multi-component mixture can have an arbitrary number of donor and acceptor association sites. The one restriction on the theory is that molecules can have at most one pair of double bonding sites. We also incorporate the effect of hydrogen bond cooperativity in cyclic double bonds. We then apply this new association theory to 2-site and 3-site models for carboxylic acids within the polar PC-SAFT equation of state. We demonstrate the accuracy of the approach by comparison to both pure and multi-component phase equilibria data. It is demonstrated that the 3-site association model gives substantially different hydrogen bonding structure than a 2-site approach. We also demonstrate that inclusion of hydrogen bond cooperativity has a substantial effect on liquid phase hydrogen bonding structure.

Bennettd1980@gmail.com

**I: Introduction**

Phase equilibria predictions of multi-component fluids are of immense industrial importance. Process development, design and optimization all rely on accurate thermodynamic models to predict fluid-phase equilibria. The challenge is to predict the phase behavior of complex mixtures using a limited amount of pure component and binary phase equilibria data. This has been particularly challenging for mixtures with associating (hydrogen bonding) components due to the directionality, limited valence and strength of the association interaction.

In the 1980's Wertheim[1–4] developed a new multi-density form of statistical mechanics which allows for the development of simple and accurate thermodynamic perturbation theories (TPT) for associating fluids. In first order perturbation theory (TPT1) each association bond in a cluster is treated independently, which allows for a general multi-component solution[5] to TPT1 where each molecule is allowed to have an arbitrary number and functionality of association sites. It is this TPT1 theory which provides the association contribution the statistical associating fluid theory (SAFT) class of EoS[6–8] as well as the Cubic Plus Association (CPA) EoS.[9]

While the generality of the TPT1 association theory allows for wide application, the restriction that each association bond must be independent results in many limitations. TPT1 cannot be used to describe ring formation[10], double bonding of molecules[11], double bonding of association sites[12] or steric hindrance[13].

Carboxylic acids are an important class of molecules for which a TPT1 treatment is inadequate due to the formation of strong double bonds. Acids form networks of hydrogen bonds in the acid rich liquid phase, but they form strong double bonds in acid dilute phases.[14] While hydrogen bond networks are described by TPT1, double bonding is a second order effect (a double bonded cluster includes a minimum of two association bond) and cannot be described with a first order treatment. Nevertheless, TPT1 has been applied to carboxylic acids with some success using 1A (single association site)[15] and 2B (single donor and single acceptor)[15,16] association schemes. The 1A scheme is more accurate for the prediction of heats of vaporization, and VLE with hydrocarbons.[15] However, as expected due to incorrect hydrogen bond stoichiometry, the 1A scheme performs poorly for mixtures of carboxylic acids and water.[17,18] To obtain a generally accurate TPT theory for carboxylic acids, the effect of double bonding needs to be included in the theory.

Sear and Jackson (SJ)[11] developed a second order perturbation theory (TPT2) which includes double bonding for a single component fluid with a 2B association scheme and double bonding. In this work we will refer to this scheme as a 2B-DB association model. Janecek and Paricaud (JP)[19–22] incorporated the SJ double bonding approach into the Perturbed Chain Statistical Associating Fluid Theory (PC-SAFT)[6] EoS. JP demonstrated a significant improvement in PC-SAFT predictions for pure component and mixture phase behavior of carboxylic acids. JP demonstrated that inclusion of the double bonding contribution allowed for the accurate representation of heats of vaporization and mixture phase equilibria with alkanes, alcohols and water.

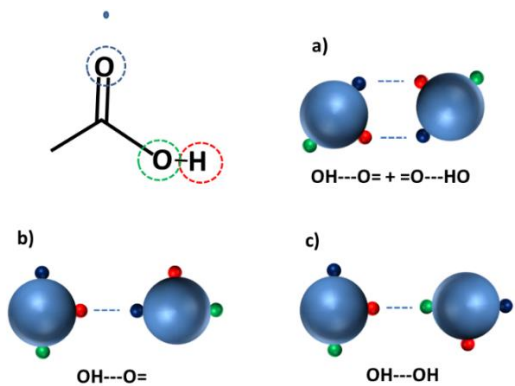

**Figure 1:** Double bonding 3-site representation of acetic acid

A drawback of both the SJ and JP approaches is a lack of generality of the derivation of the association free energy. The SJ double bonding theory restricts double bonding molecules to a 2B-DB association scheme. Figure 1 gives an example of a 3-site treatment of carboxylic acids, where each oxygen is treated as an acceptor site, with only the ketone oxygen participating in a double bond. We will refer to this approach as the 3C-DB association scheme. A significant difference between the 3C-DB and the 2B-DB approach of SJ, is that when a double bond is formed in the 2B-DB model (panel a) of Fig. 2, the double bonded molecules cannot exist in larger non-dimer clusters. The association is quenched. However, in the 3C-DB case, the double bonded molecules need not exist in a dimer. They can exist in larger associated clusters as shown in panel b) of Fig. 2. This difference can have a profound effect on hydrogen bonded structure in dense phases.

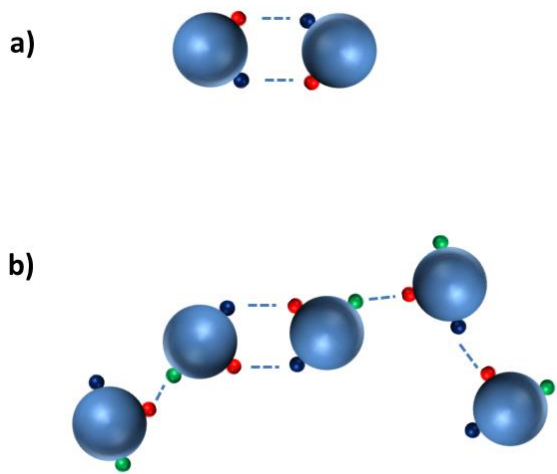

**Figure 2:** Comparison of allowed cluster classes of double bonded molecules assuming a **a)** 2-site 2B-DB and **b)** 3-site 3C-DB scheme

In this work, we remove these restrictions. We extend the dimer graph of SJ to the case of double bonding molecules with an arbitrary number of acceptor and donor association sites. Using this new double bonding contribution, we derive a TPT2 association theory which is valid for multi-component fluids with an arbitrary number of acceptor and donor association sites. The one restriction on the derivation, is that each molecule in the fluid can have at most one set of association sites which participate in cyclic double bonds.

We also include the effect of hydrogen bond cooperativity in cyclic double bonds for the first time. It is known that there is strong hydrogen bond cooperativity in the cyclic dimer.[23] That is, the hydrogen bond energy of the double bonded state is greater than twice a non-cyclic hydrogen bond. We incorporate this known physics in the foundation of the model. TPT has been shown to accurately describe the effects of hydrogen bond cooperativity.[24–26] We then apply this new TPT2 association theory to the case of 2B-DB and 3C-DB carboxylic acids as outlined in Figures 1 and 2. The association theory is then integrated with the polar PC-SAFT to obtain a complete EoS. Finally, in section V we parameterize and apply the model to predict the phase behavior of carboxylic acid mixtures using both the 2B-DB and 3C-DB association models.

## II: Thermodynamic perturbation theory

In this section we extend thermodynamic perturbation theory (TPT) to account for double bonding in a multi-component fluid.[3,4,27] We consider a mixture of $N$ molecules consisting of $n$ separate species of number density $\rho^{(k)}$. Each species contains a set of $\Gamma^{(k)} = \{A, B, C,...,G\}$ association sites, where the capitals letters represent distinct association sites. While each species can have any number and type of association sites, we restrict the theory such that only a single pair of association sites per species can participate in double bonds. Figure 1 gives a model representation of a carboxylic acid using this approach. There is one donor hydrogen and two acceptor oxygens. However, only the =O oxygen can participate in double bonds while the OH oxygen cannot.

The potential of interaction between a molecule 1 of species $k$ and a molecule 2 of species $j$ is given by

$$\phi^{(k,j)}(12) = \phi_{hs}^{(k,j)}(r_{12}) + \sum_{A \in \Gamma^{(k)}} \sum_{B \in \Gamma^{(j)}} \phi_{AB}^{(k,j)}(12) + \Delta\phi_{CDDC}^{(k,j)}(12) \qquad (1)$$

The distance between the centers of the molecules is $r_{12}$ and the notation (1) represents the position and orientation of molecule 1. The term $\varphi_{hs}^{(k,j)}(r_{12})$ is the pair potential of the spherically symmetric hard sphere reference fluid and $\varphi_{AB}^{(k,j)}$ is the potential of interaction between site $A$ on species $k$ and site $B$ on species $j$. The term $\Delta\varphi_{CDDC}^{(k,j)}(12)$ accounts for hydrogen bond cooperativity in the cyclic double hydrogen bonded structure.

The theory is developed in Wertheim's multi-density formalism.[3] In this approach each bonding state of a molecule is treated as a distinct species and assigned a density $\rho_\alpha^{(k)}$, where $\alpha$ is the set of bonded sites. Hence, $\rho_o^{(k)}$ is the monomer density of species $k$. To aid in the topological reduction from fugacity to density graphs, Wertheim defines a set of density parameters

$$\sigma_\alpha^{(k)} = \sum_{\gamma \subset \alpha} \rho_\alpha^{(k)} \qquad (2)$$

where $\sigma_o^{(k)} = \rho_o^{(k)}$ and $\sigma_\Gamma^{(k)} = \rho^{(k)}$. The total Helmholtz free energy is given by[4]

$$\frac{A - A_{hs}}{Vk_BT} = \sum_k \left( \rho^{(k)} \ln\left(\frac{\rho_o^{(k)}}{\rho^{(k)}}\right) + Q^{(k)} + \rho^{(k)} \right) - \frac{\Delta c^{(o)}}{V} \qquad (3)$$

where $A_{hs}$ is the free energy of the reference fluid, $V$ is the system volume and $T$ is the absolute temperature.

Equation (3) is mathematically exact. The fundamental graph sum $\Delta c^{(o)}$ is an infinite series of integrals which encodes all association interactions between molecules. In this work we consider two classes of association attractions. First, we allow for standard trees of association bonds in first order perturbation theory (TPT1). TPT1 neglects steric hindrance between association sites[27], hence association at one site does not hinder association at another. In double bonding molecules such as carboxylic acids, the double bonding sites will be in close proximity; hence, the neglect of steric effects in TPT1 will introduce some error. Second, we allow for the possibility of the formation of double bonds between two pairs of association sites on different molecules. Whether or not two molecules form double bonds will depend on the specific association potential of interaction. The formation of double bonds is a second order effect (requires a minimum of two association bonds) and must be treated in second order perturbation theory (TPT2).

With these considerations the sum $\Delta c^{(o)}$ is decomposed into a first order contribution $\Delta c^{(TPT1)}$ and a contribution for double bonding $\Delta c^{(DB)}$

$$\Delta c^{(o)} = \Delta c^{(TPT1)} + \Delta c^{(DB)} \qquad (4)$$

The first order contribution is obtained by summing over all contributions to $\Delta c^{(o)}$ which contain a single association bond

$$\Delta c^{(TPTI)}/V = \frac{1}{2}\sum_{k=1}^{n}\sum_{j=1}^{n}\sum_{A\in\Gamma^{(k)}}\sum_{B\in\Gamma^{(j)}}\sigma^{(k)}_{\Gamma^{(k)}-A}\sigma^{(j)}_{\Gamma^{(j)}-B}\Delta^{(k,j)}_{AB} \qquad (5)$$

where

$$\Delta^{(k,j)}_{AB} = \frac{1}{8\pi}\int f^{(k,j)}_{AB}(12)g^{(k,j)}_{hs}(r_{12})d(12) \qquad (6)$$

The association Mayer function is given by

$$f^{(k,j)}_{AB}(12) = \exp\left(-\frac{\phi^{(k,j)}_{AB}(12)}{k_b T}\right) - 1 \qquad (7)$$

and $g^{(k,j)}_{hs}(r_{12})$ is the mixture pair correlation function of the hard sphere reference system.

For the double bonding contribution we generalize the approach of Sear and Jackson[11] to the current case of mixtures, where each molecule has an arbitrary number of association sites, and there is hydrogen bond cooperativity. Again, we assume that each associating molecule has a <u>maximum</u> of 1 pair of association sites which can participate in a dimer bond. For notational simplicity this pair of sites on each molecule {C,D} which yields a dimer contribution

$$\Delta c^{(DB)}/V = \frac{1}{2}\sum_{k=1}^{n}\sum_{j=1}^{n}\sigma^{(k)}_{\Gamma^{(k)}-CD}\sigma^{(j)}_{\Gamma^{(j)}-CD}\Delta^{(k,j)}_{CDDC} \qquad (8)$$

$$\Delta^{(k,j)}_{CDDC} = \frac{1}{8\pi}\int f^{(k,j)}_{CD}(12)f^{(k,j)}_{DC}(12)g^{(k,j)}_{hs}(r_{12})\exp\left(-\frac{\Delta\phi^{(k,j)}_{CDDC}(12)}{k_b T}\right)d(12) \qquad (9)$$

In Equation (9) we have included the cooperative contribution to double bonding $\Delta\varphi^{(k,j)}_{CDDC}$ as a Boltzmann factor to further stabilize the cyclic dimer hydrogen bonds.

Equation (9) completes the definition of the graph sum $\Delta c^{(o)}$. The last term to consider in Eq. (3) is $Q^{(k)}$

$$Q^{(k)} = -\rho^{(k)} + \sum_{\substack{\gamma\subset\Gamma^{(k)} \\ \gamma\neq\emptyset}} c^{(k)}_{\gamma}\sigma^{(k)}_{\Gamma^{(k)}-\gamma} \qquad (10)$$

The functions $c_\gamma^{(k)}$ are generated from the graph sum $\Delta c^{(o)}$ according to the relation

$$c_\gamma^{(k)} = \frac{\partial}{\partial \sigma_{\Gamma-\gamma}^{(k)}} \frac{\Delta c^{(o)}}{V} ; \quad \gamma \neq \emptyset \tag{11}$$

Evaluating Eq. (11) subject to Eqns. (4), (5) and (8)

$$c_A^{(k)} = \sum_{j=1}^{n} \sum_{B \in \Gamma^{(j)}} \rho^{(j)} X_B^{(j)} \Delta_{AB}^{(k,j)} \tag{12}$$

$$c_{AB}^{(k)} = \begin{cases} \sum_{j=1}^{n} \rho^{(j)} X_{CD}^{(j)} \Delta_{CDDC}^{(k,j)} & \text{for } AB = CD \\ 0 & \text{otherwise} \end{cases} \tag{13}$$

$$c_\gamma^{(k)} = 0 \quad \text{for} \quad n(\gamma) > 2 \tag{14}$$

It is the non-independence of association sites $c_{CD}^{(k)} \neq 0$ which makes the theory second order. A key quantity in TPT1 is the fraction molecules not bonded at site A: $X_A^{(k)} = \sigma_{\Gamma^{(k)}-A}^{(k)}/\rho$. In this second order theory we will also require the fraction of molecules not bonded at both sites $C$ and $D$: $X_{CD}^{(k)} = \sigma_{\Gamma^{(k)}-CD}^{(k)}/\rho$.

With the current assumption of only a single pair of second order (double bonding) sites, the theory will have similar structure to that of Marshall and Chapman (MC)[28]. Generalizing the MC solution to a multi-component mixture we obtain

$$X_S^{(k)} = \begin{cases} \dfrac{1}{1+c_S^{(k)}} & \text{for } S \neq C \text{ or } D \\ \left(1+c_L^{(k)}\right) X_{CD}^{(k)} & \text{otherwise} \end{cases} \tag{15}$$

In Eq. (15) when $S = C$, $L = D$ and when $S = D$, $L = C$. The fractions $X_{CD}^{(k)}$ are given by

$$X_{CD}^{(k)} = \frac{1}{c_{CD}^{(k)} + \left(1 + c_C^{(k)}\right)\left(1 + c_D^{(k)}\right)} \tag{16}$$

The monomer fraction is found to be

$$X_o^{(k)} = \frac{X_{CD}^{(k)}}{X_C^{(k)} X_D^{(k)}} \prod_{A \in \Gamma^{(k)}} X_A^{(k)} \tag{17}$$

From these results Eq. (10) can be evaluated as

$$Q^{(k)} / \rho^{(k)} = \sum_{A \in \Gamma^{(k)}} \left(1 - X_A^{(k)}\right) + \frac{X_C^{(k)} X_D^{(k)}}{X_{CD}^{(k)}} - 2 \tag{18}$$

Combining these results, we simplify the free energy in Eq. (3) to

$$\frac{A - A_{hs}}{V k_B T} = \sum_{k=1}^n \rho^{(k)} \sum_{A \in \Gamma^{(k)}} \left(\ln X_A^{(k)} - \frac{X_A^{(k)}}{2} + \frac{1}{2}\right) + \sum_{k=1}^n \rho^{(k)} \left(\ln\left(\frac{X_{CD}^{(k)}}{X_C^{(k)} X_D^{(k)}}\right) - \frac{\chi_{DB}^{(k)}}{2}\right) \tag{19}$$

where the fraction of molecules of species $k$ which are actively participating in a cyclic double bond (irrespective of the bonding state of the non $\{C,D\}$ sites) is calculated as

$$\chi_{DB}^{(k)} = \sum_{j=1}^n \rho^{(j)} X_{CD}^{(k)} X_{CD}^{(j)} \Delta_{CDDC}^{(k,j)} \tag{20}$$

To maintain consistency with the PC-SAFT[6] EoS, equation (6) is evaluated as

$$\Delta_{AB}^{(k,j)} = \sigma_{kj}^3 \kappa_{AB}^{(k,j)} \left(\exp\left(\frac{\varepsilon_{AB}^{(k,j)}}{k_b T}\right) - 1\right) g_{hs}^{(k,j)} \tag{21}$$

where $\sigma_{kj}$ is the cross-species diameter, $\kappa_{AB}^{(k,j)}$ is the bond volume (dimensionless), $\varepsilon_{AB}^{(k,j)}$ is the association energy and $g_{hs}^{(k,j)}$ is the contact value of hard sphere pair correlation function. Equation (21) can be developed by assuming both the association energy and pair correlation function remain constant within the bond volume. The mixture quantities are evaluated with the following combining rules[6]

$$\sigma_{kj}^3 \kappa_{AB}^{(k,j)} = \sqrt{\sigma_{kk}^3 \kappa_{AB}^{(k,k)} \sigma_{jj}^3 \kappa_{AB}^{(j,j)}}$$

(22)

$$\varepsilon_{AB}^{(k,j)} = \frac{\varepsilon_{AB}^{(k,k)} + \varepsilon_{AB}^{(j,j)}}{2}$$

In approximating Eq. (9), we maintain consistency with Eq. (21) by assuming both the association energy and the pair correlation function remain constant throughout the bond volume. Hence for a double bonded molecule the total double bonding energy is given by $2\varepsilon_{CD}^{(k,j)} + \Delta\varepsilon_{CDDC}^{(k,j)}$. With these assumptions we evaluate Eq. (9) as

$$\Delta_{DCCD}^{(k,j)} = \sigma_{kj}^3 \Gamma_{DB}^{(k,j)} \left( \exp\left(\frac{\varepsilon_{CD}^{(k,j)}}{k_b T}\right) - 1 \right)^2 \exp\left(\frac{\Delta\varepsilon_{CDDC}^{(k,j)}}{k_b T}\right) g_{hs}^{(k,j)}$$

(23)

where $\Gamma_{DB}^{(k,j)}$ is the double bond volume which is proportional to the number of double bonded states. The cross species double bond volume is evaluated with the combining rule

$$\sigma_{kj}^3 \Gamma_{DB}^{(k,j)} = \sqrt{\sigma_{kk}^3 \Gamma_{DB}^{(k,k)} \sigma_{jj}^3 \Gamma_{DB}^{(j,j)}}$$

(24)

and the cooperative contribution $\Delta\varepsilon_{CDDC}^{(k,j)}$ is evaluated with an arithmetic mean combining rule in accordance with Eq. (22).

In this work we will follow the simplified approach[29] of evaluating the mixture contact pair correlation function in Eqns. (21) and (23) with the pure component Carnahan and Starling[30] EoS evaluated with the mixture packing fraction $g_{hs}^{(k,j)} \sim g_{hs} = (1 - \eta/2) / (1 - \eta)^3$. It has been demonstrated[29] that this approach yields an equally capable equation of state as compared to the un-simplified form. In the current work, this assumption allows for the derivation of a comparatively simple form of the association contribution to the chemical potential in Eq. (25)

$$\frac{\mu^{(k)} - \mu_{hs}^{(k)}}{k_B T} = \sum_{A \in \Gamma^{(k)}} \ln X_A^{(k)} - \frac{1}{2} \frac{\partial \ln g_{hs}}{\partial \rho^{(k)}} \sum_{j=1}^{n} \sum_{A \in \Gamma^{(j)}} \left(1 - X_A^{(j)}\right) \rho^{(j)} + \ln\left(\frac{X_{CD}^{(k)}}{X_C^{(k)} X_D^{(k)}}\right) + \cdots +$$
$$\frac{\partial \ln g_{hs}}{\partial \rho^{(k)}} \sum_{j=1}^{n} \left(1 - \frac{X_C^{(j)} X_D^{(j)}}{X_{CD}^{(j)}} - \frac{\chi_{DB}^{(j)}}{2}\right) \rho^{(j)}$$

(25)

The theory developed in this section is general for an arbitrary number of components, with an arbitrary number of acceptor and donor association sites. The one restriction is that each molecule can have at most one pair of association sites which participate in cyclic double bonds. In section III the theory is specialized to carboxylic acids using both the 3C-DB and 2B-DB association models.

**III: Application to carboxylic acids**

In this section we specialize the theory developed in section II to the case of mixtures which contain carboxylic acids with the 3C-DB association model outlined in Fig. 1. Site $H$ (red) is the hydrogen bond donor site, $O_1$ is the carbonyl oxygen acceptor site (blue) and finally $O_2$ is the acceptor site associated with the -OH oxygen (green). Both $H$ and $O_1$ can participate in cyclic double bonds, while $O_2$ cannot. From Eqns. (12) – (16) we obtain the needed bonding fractions

$$X_{O_2}^{(k)} = \frac{1}{1+\sum_{j=1}^{n}\sum_{B\in\Gamma^{(j)}} \rho^{(j)} X_B^{(j)} \Delta_{O_2 B}^{(k,j)}} \quad (26)$$

$$X_{O_1}^{(k)} = \left(1+\sum_{j=1}^{n}\sum_{B\in\Gamma^{(j)}} \rho^{(j)} X_B^{(j)} \Delta_{HB}^{(k,j)}\right) X_{O_1 H}^{(k)} \quad (27)$$

$$X_H^{(k)} = \left(1+\sum_{j=1}^{n}\sum_{B\in\Gamma^{(j)}} \rho^{(j)} X_B^{(j)} \Delta_{O_1 B}^{(k,j)}\right) X_{O_1 H}^{(k)} \quad (28)$$

$$X_{O_1 H}^{(k)} = \frac{1}{\sum_{j=1}^{n} \rho^{(j)} X_{O_1 H}^{(j)} \Delta_{O_1 H H O_1}^{(k,j)} + \left(1+\sum_{j=1}^{n}\sum_{B\in\Gamma^{(j)}} \rho^{(j)} X_B^{(j)} \Delta_{O_1 B}^{(k,j)}\right)\left(1+\sum_{j=1}^{n}\sum_{B\in\Gamma^{(j)}} \rho^{(j)} X_B^{(j)} \Delta_{HB}^{(k,j)}\right)} \quad (29)$$

For associating molecules which do not participate in double bonds, the necessary bonding fractions are that of the standard first order form given by Eqn. (26). Of course, all bonding fractions are coupled, requiring the simultaneous solution of a system of non-linear equations. We solve the set of equations using a Newton's method with an analytical Jacobian.

For clarity, we now derive the fraction of molecules which participate in cyclic double bonds given by Eq. (20). Within Wertheim's multi-density formalism the density of molecules bonded as the set of sites $\gamma$ is given by[3]

$$\frac{\rho_\gamma^{(k)}}{\rho_o^{(k)}} = \sum_{P(\gamma)=\{\tau\}} \prod_\tau c_\tau^{(k)} \qquad (30)$$

where $P(\gamma)$ is the partition of the set $\gamma$ into non-empty subsets. For the current 3C-DB model, the density of molecules which are bonded at both sites $H$ and $O_1$, but not bonded at $O_2$ is

$$\frac{\rho_{O_1 H}^{(k)}}{\rho_o^{(k)}} = c_{O_1}^{(k)} c_H^{(k)} + c_{O_1 H}^{(k)} \qquad (31)$$

The product $c_{O_1}^{(k)} c_H^{(k)}$ accounts for the density of molecules which are bonded at both sites, but do not participate in cyclic double bonds, while $c_{O_1 H}^{(k)}$ accounts for molecules which are doubly bonded. Similarly, the density of fully hydrogen bonded carboxylic acids is given by

$$\frac{\rho_{O_1 H O_2}^{(k)}}{\rho_o^{(k)}} = c_{O_1}^{(k)} c_H^{(k)} c_{O_2}^{(k)} + c_{O_2}^{(k)} c_{O_1 H}^{(k)} \qquad (32)$$

where the product $c_{O_2}^{(k)} c_{O_1 H}^{(k)}$ accounts for the density of molecules which are fully bonded, with $H$ and $O_1$ participating in a cyclic double bond. The fraction of double bonded molecules is then given by the sum of double bonded molecules which are both unbonded (31) and bonded (32) at $O_2$

$$\chi_{DB}^{(k)} = X_o^{(k)} \left(1 + c_{O_2}^{(k)}\right) c_{O_1 H}^{(k)} \qquad (33)$$

The fraction of molecules which are unbonded is given by Eq. (17)

$$X_o^{(k)} = X_{O_1 H}^{(k)} X_{O_2}^{(k)} \qquad (34)$$

Combining Eqns. (13), (15), (33) - (34) we obtain

$$\chi_{DB}^{(k)} = \sum_{j=1}^{n} \rho^{(j)} X_{O_1H}^{(k)} X_{O_1H}^{(j)} \Delta_{O_1HHO_1}^{(k,j)} \tag{35}$$

Unlike previous theories which include double bonding, the fraction $\chi_{DB}^{(k)}$ includes molecules which are both bonded and unbonded at the third site $O_2$. From Eqns. (34) - (35) we can then calculate the fraction of molecules which are hydrogen bonded, but do not participate in a cyclic double bond

$$\chi_{NO:DB}^{(k)} = 1 - \chi_{DB}^{(k)} - X_o^{(k)} \tag{36}$$

In general, the association energies $\varepsilon_{O_1H}, \varepsilon_{O_2H}$ and bond volumes $\kappa_{O_1H}, \kappa_{O_2H}$ need not be equal. This is particularly true of the bond volumes, as the OH---O= bond has less steric hindrance than the OH---OH bond. However, it would difficult to regress the parameters in a meaningful way. Instead we assume that the association energies and bond volumes are equal

$$\varepsilon_{OH} = \varepsilon_{O_2H} = \varepsilon_{O_1H}$$
$$\kappa_{OH} = \kappa_{O_1H} = \kappa_{O_2H} \tag{37}$$

We must also estimate the cooperative contribution observed in cyclic double bonds $\Delta\varepsilon_{HOOH}$. We do not adjust this quantity to data, but instead appeal to *ab initio* calculations[23] which have shown that the cooperative contribution to the dimer energy of acetic acid is about 25% - 50% of the energy of single hydrogen bond. We take the lower end of this scale and assume the following relation for carboxylic acids

$$\Delta\varepsilon_{HOOH} = \frac{\varepsilon_{OH}}{4} \tag{38}$$

With Eqns. (37) - (38) the association contribution to the free energy for carboxylic acids is fully described by 3 parameters $\varepsilon_{OH}$, $\kappa_{OH}$ and $\Gamma_{DB}$.

For the 2B-DB model outlined in panel a) of Fig. 2, there is only a single acceptor oxygen site which we label $O_1$. Hence, for this case, Eq. (26) is neglected and the carboxylic acid association is described by Eqns. (27) – (29).

**IV: Incorporation into polar PC-SAFT**

The association contribution alone is not sufficient to define the full equation of state. For this, we incorporate the association theory into the wider polar PC-SAFT approach. The total excess (residual) Helmholtz free energy is defined as

$$a_{ex} = \frac{A_{ex}}{Nk_BT} = a_{hc} + a_{at} + a_{dp} + a_{as} \tag{39}$$

where $a_{hc}$ is the excess free energy of the hard chain reference fluid which we evaluate with the simplified approach of von Solmes et al.[29]. The contribution to the theory due to spherically symmetric attractions $a_{at}$ is evaluated with the PC-SAFT approach[31] of Gross and Sadowski. The contribution to the free energy due to long range dipolar interactions $a_{dp}$ is evaluated with the Jog and Chapman[32] polar contribution. Finally, the association contribution to the free energy $a_{as}$ is evaluated with Eq. (19).

**Section V: Parameterization and model results**

In this section we discuss parameterization of the model and compare to experimental vapor-liquid equilibria (VLE) data. As discussed in section III, hydrogen bonding molecules which do not participate in cyclic double bonds can have any number of donor and acceptor sites. However, carboxylic acids which participate in cyclic double bonds are restricted to having either the 2B-DB or 3C-DB association schemes. The number and type of association sites are imposed on the model and are not treated as variables. The remaining parameters of the model are illustrated in Fig. 3.

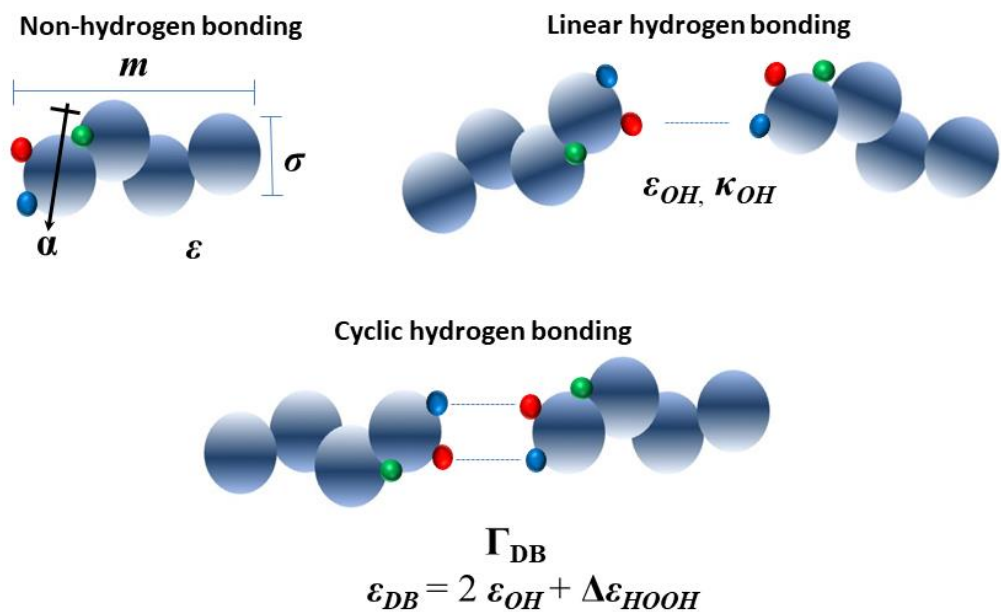

**Figure 3:** Illustration of model parameters classes

Non-polar molecules are described by the segment diameter $\sigma$, chain length $m$, and square well attractive energy $\varepsilon$. The long range polar contribution to the theory is parameterized by the dipole moment $\mu$ and fraction of polar segments $x_p$. In the Jog and Chapman polar contribution the parameters $m$, $\mu$ and $x_p$ enter the theory as the product $\alpha$ which we term the polar strength

$$\alpha = m x_p \mu^2 \qquad (40)$$

As usual, model parameters are adjusted to reproduce pure component vapor pressure $P_{sat}$ and saturated liquid density $\rho$ data. For vapor pressures we only include data for which $P_{sat} > 1$ torr, and for saturated liquid densities we excluded temperatures which where greater than 90% of the critical temperature. In addition to this data, we also include heat of vaporization $\Delta H_{vap}$ data to enable the optimum choice of the double bond volume $\Gamma_{DB}$. $\Delta H_{vap}$ is a good measure of vapor phase dimerization[19]. $\Delta H_{vap}$ data was typically included in the temperature range 300 K $< T <$ 530 K. Finally, for acetic acid only, we also include VLE data with heptane to guide the optimal choice of the polar strength $\alpha$. For the remaining carboxylic acids we do not include binary data, but we enforce a constant $\alpha$ across the homologous series.

| Compound | m | σ (A) | ε / $k_b$ | $ε_{AB}$ / $k_b$ | $κ_{AB}$ | $Γ_{DB}$ 10$^4$ | α (D$^2$) | AAD $P_{sat}$ | ρ | $\Delta H_{vap}$ |
|---|---|---|---|---|---|---|---|---|---|---|
| Ethanoic acid | 2.092 | 3.303 | 200.58 | 2903.69 | 0.0288 | 5.897 | 2.138 | 2.20% | 0.15% | 1.36% |
| Propanoic acid | 2.481 | 3.412 | 212.46 | 2810.80 | 0.0238 | 8.078 | 2.138 | 2.45% | 0.26% | 2.22% |
| Butanoic acid | 2.902 | 3.480 | 218.33 | 2850.64 | 0.0184 | 5.493 | 2.138 | 1.58% | 0.50% | 1.55% |
| Pentanoic acid | 3.393 | 3.497 | 217.79 | 2850.67 | 0.0224 | 6.872 | 2.138 | 2.63% | 0.47% | 1.89% |

**Table 1:** Model parameters and average absolute deviations (AAD) from regression data for the 3C-DB association model. Parameter $\alpha$ is in units Debeye squared.

| Compound | m | σ (A) | ε / kb | $ε_{AB}$ / $k_b$ | $κ_{AB}$ | $Γ_{DB}$ 10$^4$ | α (D$^2$) | AAD $P_{sat}$ | ρ | $\Delta H_{vap}$ |
|---|---|---|---|---|---|---|---|---|---|---|
| Ethanoic acid | 2.401 | 3.162 | 195.60 | 2934.24 | 0.1121 | 4.814 | 1.890 | 0.26% | 0.45% | 1.10% |
| Propanoic acid | 2.914 | 3.266 | 201.51 | 2814.34 | 0.1115 | 5.243 | 1.890 | 0.47% | 2.50% | 6.00% |
| Butanoic acid | 3.447 | 3.281 | 202.77 | 2913.59 | 0.1010 | 4.020 | 1.890 | 0.65% | 0.50% | 7.10% |
| Pentanoic acid | 3.919 | 3.318 | 204.62 | 2962.73 | 0.1075 | 5.823 | 1.890 | 1.87% | 0.53% | 2.70% |

**Table 2:** Same as Table 1 except for 2B-DB association model

The parameters and average absolute deviations (AAD) can be found in table 1 for the 3C-DB model and table 2 for the 2B-DB model. Both association models accurately describe the pure component properties of the carboxylic acids. Of particular interest is the good representation of the $\Delta H_{vap}$ data. As discussed by Janecek and Paricaud (JP)[19], standard PC-SAFT approaches which do not include double bonding cannot represent this quantity. As this has been extensively discussed by JP[19], we forgo any further analysis here.

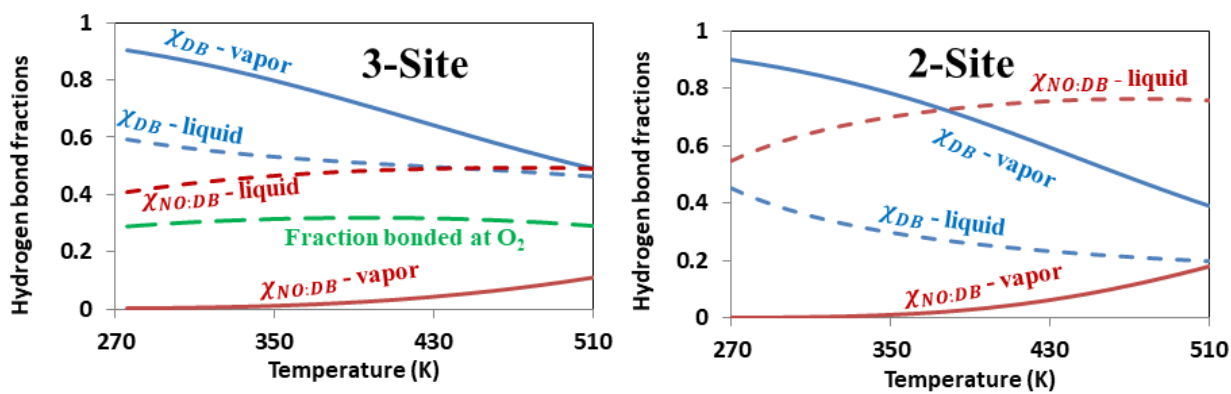

**Figure 4:** Hydrogen bonding fractions for coexisting vapor and liquid phases of acetic acid. $\chi_{DB}$ gives the fraction of molecules which are double bonded and $\chi_{NO:DB}$ gives the fraction of molecules which are hydrogen bonded, but not double bonded. Left panel gives the 3C-DB association model results, while the right panel gives the 2B-DB site results. Long dashed green curve on left panel gives fraction of 3C-DB acids bonded at the oxygen site $O_2$.

The 3C-DB model and the 2B-DB model give markedly different hydrogen bonding structure. This can be seen in Fig. 4 which plots the fraction of molecules which are participating in double bonds $\chi_{DB}$ given by Eq. (35) and the fraction of molecules which are hydrogen bonded but do not participate in a double bond $\chi_{NO:DB}$ given by Eq. (36). These results are for coexisting vapor and liquid phases of acetic acid. Both approaches predict that the vapor phase is dominated by doubly bonded molecules at low pressure. However, the 3C-DB case predicts much more double bonding in the liquid phase than the 2B-DB case. The difference is entropic in nature. When the 2B-DB acid participates in a double bond, it becomes hydrogen bond saturated as shown in panel a) of Fig. 2. However, the additional hydrogen bond acceptor site in the 3C-DB model allows for the doubly bonded acid to be incorporated in larger clusters as in panel b) of Fig. 2. The long dashed green curve in the left panel of Fig. 4 gives the fraction of molecules bonded at the non-double bonding site $O_2$. This results in a lower entropic penalty to double bond formation, and a correspondingly higher degree of double bonding in the liquid.

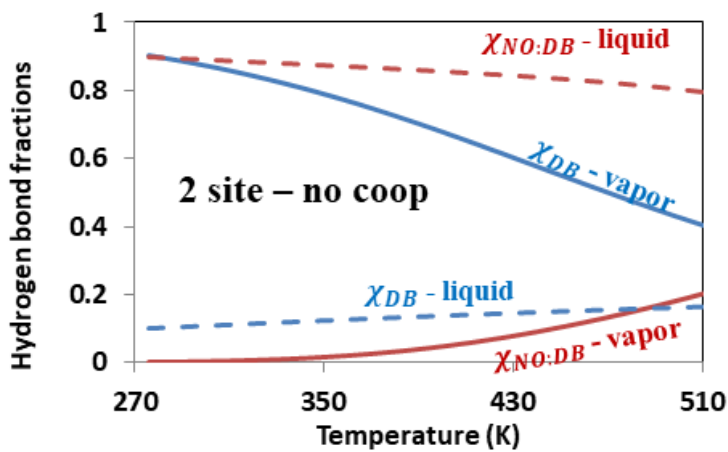

**Figure 5:** Same as right panel of Fig. 4 except using theory and parameter set which do not include hydrogen bond cooperativity.

Another interesting feature of both approaches in Fig. 4 is that the liquid phase $\chi_{DB}$ decreases with increasing temperature. This is in contrast to the results of JP[19] which showed exactly the opposite dependence, and less double bonding overall in the liquid state than both the 3C-DB and 2B-DB models presented here. As we shall now demonstrate, the source of this discrepancy is that we include the effect of hydrogen bond cooperativity dictated through Eq. (38). To demonstrate this, we refit the model parameters for acetic acid in the exact same manner as the 2B-DB parameters in table 2, with the one exception, that we set the cooperative contribution in Eq. (38) is set to zero. That is, we do not include hydrogen bond cooperativity. Figure 5 presents the hydrogen bond fraction calculated from the resulting parameter set. As can be seen, the hydrogen bonded structure is markedly different than the right panel of Fig. 4, which does include hydrogen bond cooperativity. The results in Fig. 5 are in qualitative agreement with the results of JP. Comparing the 2B-DB approaches from Fig. 4 and Fig.5 we see that the inclusion of hydrogen bond cooperativity increases the fraction of double bonded molecules in the liquid phase, and results in a decreasing fraction of liquid phase double bonds with increasing temperature.

The recent neutron diffraction data and modelling of Imberti and Bowron (IB)[14] suggest that the liquid structure of acetic acid is dominated by linear chains of hydrogen bonds between the carbonyl oxygen =O and the hydroxide hydrogen -OH. These linear chains are further stabilized by numerous weak hydrogen bonds between a methyl hydrogen -CH and the carbonyl oxygen. The effect of these weak -CH hydrogen bonds has not been incorporated into the current model. However, IB empirically adjusted a classical intermolecular pairwise additive potential using Monte Carlo simulations to reproduce the neutron diffraction data. The adjusted pair potential which was used to calculate the fluid structure was composed of a Lennard Jones core with Coulomb charges. The hydrogen bond cooperativity observed in cyclic dimers of carboxylic acids[23] is an inherently quantum effect which cannot be reproduced using classical Coulomb charges. The results of Figs. 4 and 5 demonstrate the sensitivity of the model to the cooperativity in cyclic hydrogen bonds.

Now we focus our attention on the vapor-liquid equilibria (VLE) of binary mixtures containing carboxylic acids. It was shown by JP[19], that the inclusion of double bonding significantly improves the description of binary VLE of carboxylic acids with alkanes as compared to PC-SAFT approaches which do not include double bonding. We do not repeat this analysis here, instead focusing on the differences between the approach developed in this work, and the model results of JP. Both the 3C-DB and 2B-DB association theories developed in this work give nearly identical predictions for the binary VLE with hydrocarbons. Hence, in what follows, we compare the 3C-DB model to the predictions using the model of JP.

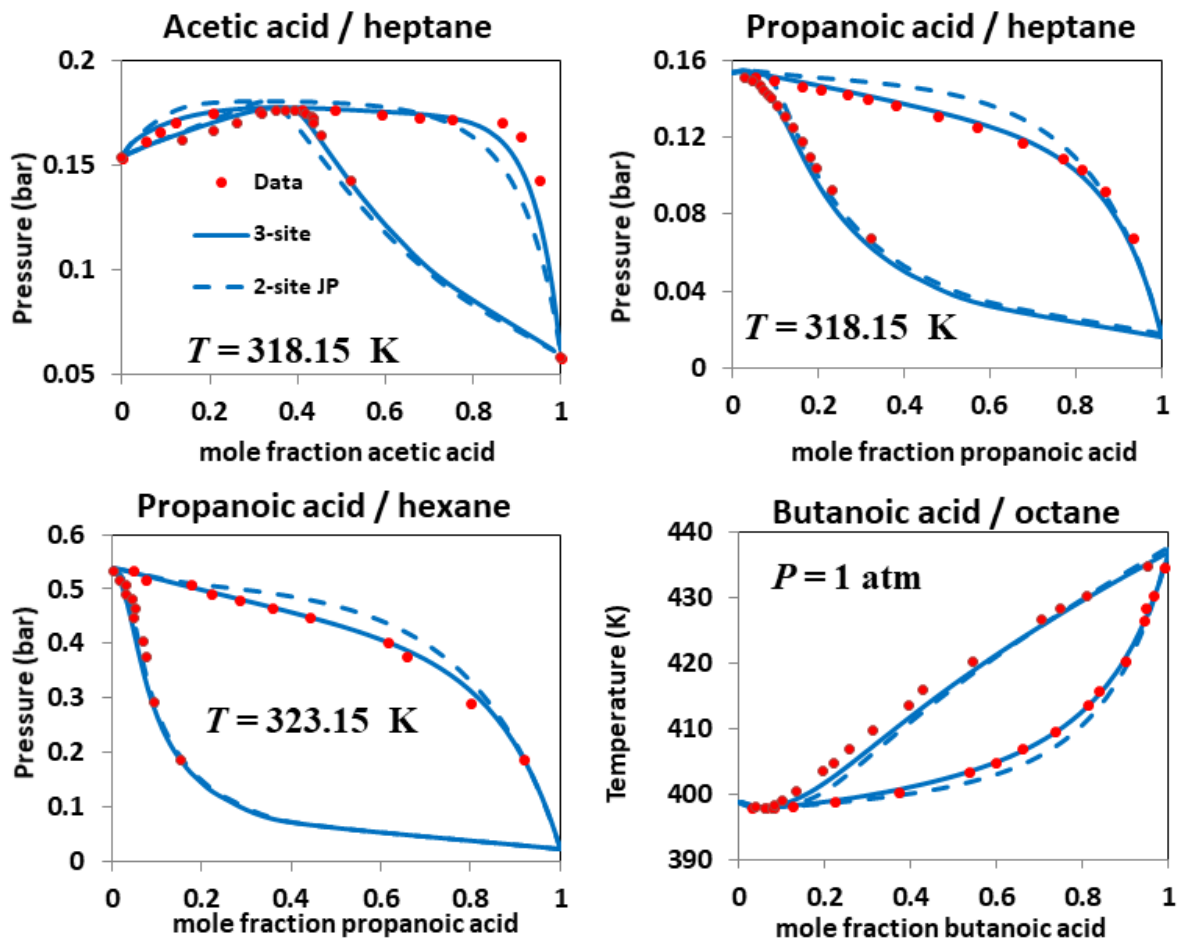

**Figure 6:** Comparison of model predictions (curves) to experimental data (circles) for the binary VLE of carboxylic acids with alkanes. Solid curves give the three 3C-DB model results while dashed curve gives model results using the 2B-DB model and parameters of JP[19] (with the simplified hard chain reference discussed in section IV). Data references: acetic acid / heptane[33], propanoic acid / heptane[34], propanoic acid / hexane[35], butanoic acid / octane[36]

Figure 6 compares model predictions for the 3C-DB and JP models to experimental data for the VLE of carboxylic acids and alkanes. For all mixture calculations using the JP model and parameters, we employ the simplified hard chain contribution[29] outlined in section IV. As compared to the JP model, the 3C-DB model gives improved predictions for the binary VLE of carboxylic acids with alkanes. However, this is not a result of the association treatment, but instead the fact that within the 3C-DB approach we have included the long ranged polar contribution to the free energy.

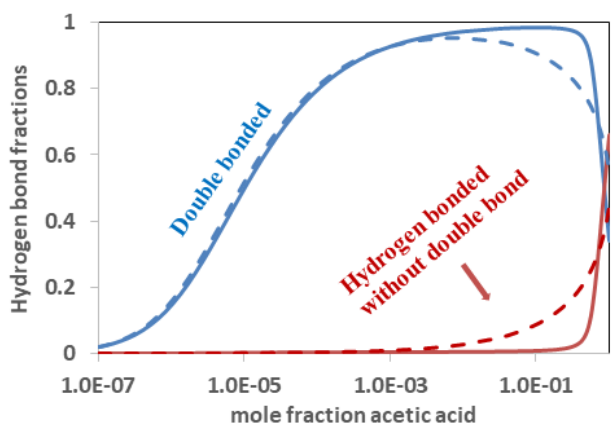

**Figure 7:** Model predictions for the hydrogen bonded structure of acetic acid in an acetic acid / heptane mixture (liquid) at $T = 318.15$ K. Blue curves give the fraction of acetic acid molecules which are double bonded $\chi_{DB}$ and red curves give the fraction of acetic acid molecules which are hydrogen bonded, but not double bonded $\chi_{NO:DB}$. Solid curves give 2B-DB results and dashed curves give 3C-DB results

As stated above, the 3C-DB and 2B-DB models developed in this work give very similar phase equilibria predictions for the VLE with hydrocarbons. However, as with pure acetic acid, the hydrogen bonded clusters predicted by the approaches can be quite different. Figure 7 shows model predictions of the hydrogen bonding clusters in the coexisting liquid phase of an acetic acid / heptane mixture at $T = 318.15$ K using both the 3C-DB and 2B-DB models developed here. The predictions correspond to the phase diagram in the top left panel of Fig. 6. As can be seen, even at very low mole fractions of acetic acid in the liquid phase, there is strong dimerization. In these dilute acetic acid liquids, the 3C-DB and 2B-DB acid models give similar predictions of hydrogen bonded structure. Increasing the mole fraction of acetic acid above order $10^{-3}$, the predictions of the two models deviate. The 3C-DB model shows a faster initial decrease in double bonded molecules as compared to the 2B-DB case. This is a result of the availability of the non-double bonding $O_2$ receptor site. Eventually the fraction of double bonding molecules in the 2B-DB model makes a sharp decrease, approaching the pure component limit dictated in Fig. 4.

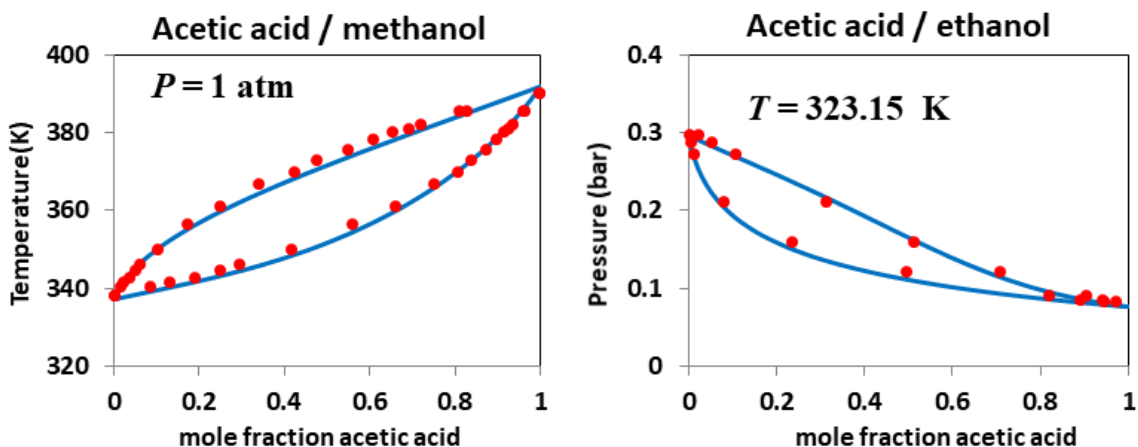

**Figure 8:** Model predictions (curves) versus data (symbols) for the VLE of acetic acid with methanol (left)[37] and ethanol (right)[38]

Now we consider cross-associating mixtures. Fouad et al.[39] showed that a 3-site (2 acceptor, and one donor) polar PC-SAFT model for alcohols was able to accurately predict the binary VLE of alcohol / water mixtures. Figure 8 compares model predictions to experimental data for VLE of acetic acid with methanol and ethanol. Acetic acid is treated with the 3C-DB model, and the alcohols are treated using the 3-site parameters of Fouad et al.[39]. As can be seen, the theoretical predictions are in good agreement with the experimental data.

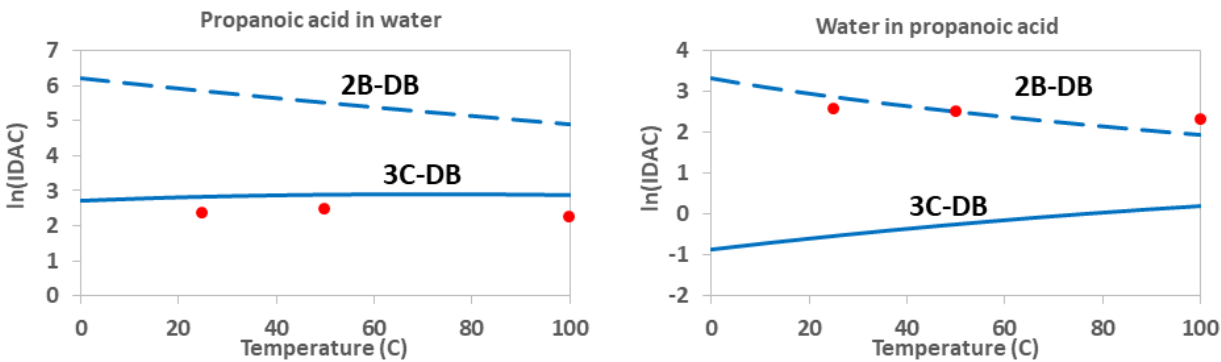

**Figure 9:** Comparison of model predictions to experimental data[40] for the natural log the IDAC for propanoic acid in water (left) and water in propanoic acid (right). Solid curve gives 3C-DB model results and the dashed curve gives 2B-DB results

To accurately describe the phase equilibria of carboxylic acid / water binary mixtures a binary interaction parameter $k_{ij}$ must be adjusted. However, a significant amount of model insight can be gained (without adjustment) by considering theory predictions of activity coefficients at infinite dilution (IDAC). Figure 9 compares model predictions of the natural log of the IDAC of propanoic acid in water (left panel) and of water in propanoic acid (right panel) to experimental data using both the 3C-DB and 2B-DB association schemes. For water we use the 4-site polar PC-SAFT parameters of Fouad *et al.*[39] For the case of propanoic acid in water, the 3C-DB scheme gives a good representation of the data, while the 2B-DB model substantially overpredicts the IDAC. This behavior is easily explicable in terms of the number of association sites. The 2B-DB model incorrectly assumes that each carboxylic acid can participate in at most two hydrogen bonds. This under-predicts the degree of solvation of the carboxylic acid, resulting in a substantial overprediction of the IDAC. On the other hand, the 3C-DB model allows for an additional association bond at the $O_2$ receptor site. This results in further solvation and a resulting decrease in the IDAC.

As can be seen in the right panel of Fig. 9, the 2B-DB model accurately predicts the IDAC of water in propanoic acid while the 3B-DB model substantially under-predicts this quantity. That is, the 3C-DB model substantially over-predicts the attractions between a lone water molecule in a pure propanoic acid phase. This is likely the result of an underlying hydrogen bonding structure of the propanoic acid phase which restricts access to the additional $O_2$ acceptor site (hydroxyl oxygen).

The neutron diffraction data and modelling of Imberti and Bowron[14] predicted that, for a pure acetic acid liquid phase, hydrogen bonds between a hydroxyl oxygen ($O_2$) and hydroxyl hydrogen ($H$) are rare. These are instead favored by weaker but more numerous CH-O hydrogen bonds. This level of structural cooperativity in the hydrogen bonding of carboxylic acids cannot be reproduced in a second order perturbation theory. Hence, the 3C-DB model predicts an overly accessible $O_2$ receptor site in a pure propanoic acid phase, and a corresponding under-prediction of the IDAC of water in propanoic acid. As the 2B-DB model does not possess this additional receptor site, it is naturally unavailable for bonding.

## VI: Summary

We have developed the first general multi-component solution to Wertheim's TPT2 for double bonding, where each species can have an arbitrary number of association sites. The one restriction on the theory is that each species can have at most one set of double bonding sites. We have also allowed for cooperative effects in cyclic double bonds. The new theory was applied to 2-site and 3 site models of carboxylic acids. Both approaches were shown to accurately represent both pure component and mixture phase equilibria. While both the 3C-DB and 2B-DB association models yields similar phase equilibrium predictions, they differ substantially in their predictions of hydrogen bonding structure. In addition, we have shown that the inclusion of hydrogen bond cooperativity has a substantial effect on liquid phase hydrogen bond structure.